\documentclass[12pt]{iopart}

\usepackage{tabularx}
\usepackage{array}
\usepackage{graphicx}
\usepackage{amssymb,amsfonts,amscd, xcolor, cite}
\usepackage{here}
\usepackage{url}
\usepackage{multicol}
\usepackage{abstract}
\usepackage{textcomp}
\usepackage{xcolor}

\begin{document}
\title{A Review of SRAM-based Compute-in-Memory Circuits}
\author{Kentaro Yoshioka*, Shimpei Ando, Satomi Miyagi, \\Yung-Chin Chen, Wenlun Zhang}
\address{Keio University Department of Electrical and Electronics Engineering Yagami Campus
3-14-1 Hiyoshi, Kohoku-ku, Yokohama, Kanagawa 223-8522, Japan}
\ead{kyoshioka47@keio.jp}
\vspace{10pt}
\begin{indented}
\item[]October 2024 
\end{indented}


\begin{abstract}
This paper presents a tutorial and review of SRAM-based Compute-in-Memory (CIM) circuits, with a focus on both Digital CIM (DCIM) and Analog CIM (ACIM) implementations. We explore the fundamental concepts, architectures, and operational principles of CIM technology. 

The review compares DCIM and ACIM approaches, examining their respective advantages and challenges. DCIM offers high computational precision and process scaling benefits, while ACIM provides superior power and area efficiency, particularly for medium-precision applications. We analyze various ACIM implementations, including current-based, time-based, and charge-based approaches, with a detailed look at charge-based ACIMs. The paper also discusses emerging hybrid CIM architectures that combine DCIM and ACIM to leverage the strengths of both approaches. 

\end{abstract}

\section{Introduction}

Recent advancements in Artificial Intelligence (AI) and Machine Learning (ML) have led to a dramatic increase in computational demands, particularly in the field of Deep Neural Networks (DNNs). The evolution of DNNs, from the breakthrough of AlexNet in 2012~\cite{krizhevsky2012imagenet} to modern architectures like ResNet~\cite{he2016resnet} and Transformers~\cite{vaswani2017attention}, has resulted in exponential growth in model complexity and scale. While these advancements have enabled remarkable achievements in various applications, including surpassing human-level performance in image recognition~\cite{he2015delving} and demonstrating impressive language understanding capabilities~\cite{brown2020language}, they have also pushed traditional computing architectures to their limits.

Traditional von Neumann architectures, which physically separate processors and memory, have become a significant bottleneck in processing AI/ML workloads~\cite{horowitz20141}. Specifically, the memory wall problem limits performance and power efficiency due to data transfer between processors and memory. Moreover, while GPUs are \textit{defacto} standard for DNN computing hardware, there is an increasing demand for improved power efficiency in AI/ML processing across all scenarios, from edge devices to data centers.
To address these challenges, researchers are exploring new computing paradigms, with Compute-In-Memory (CIM) technology \cite{sebastian2020memory, gonugondla2020fundamental, verma2019memory,valavi201964, yao2023fully, yoshioka202434, yoshioka2024818, ando24ssdm, jia2020programmable, jia202115, lee2021fully, lee2022low, chen2024pico, wu202228nmtime, dong202015current, houshmand2022diana, chen2022hybrid, chen2023osa, yuan202434hybrid, guo202434hybrid, chih202116, zhang2024pacim, tu2022trancim, tu2023multcim, fujiwara202434, khwa202434, shih202420digital} gaining particular attention. CIM aims to minimize data movement and significantly enhance parallel processing capabilities and power efficiency by physically integrating memory and computational units.

Compared to the previous SSDM2024 Extended Abstract \cite{yoshioka24ssdm}, this paper aims to provide a comprehensive review on recent advancements in the field of CIM macros, analyzing various DCIM, ACIM and hybrid architectures and their implementations. SRAM is well-suited for CIM macro implementation due to its high-speed access, low power consumption, and widespread use as on-chip memory in CMOS processes, requiring no additional process options and thus being cost-effective. 
We will particularly emphasize the comparative analysis of Digital CIM (DCIM) \cite{chih202116, zhang2024pacim, tu2022trancim, tu2023multcim, fujiwara202434, khwa202434, shih202420digital} and Analog CIM (ACIM) \cite{gonugondla2020fundamental, valavi201964, yao2023fully, yoshioka202434, yoshioka2024818, ando24ssdm, lee2021fully, jia2020programmable, jia202115, lee2022low, wu202228nmtime, dong202015current, chen2024pico}, examine the advantages and challenges of ACIM approaches, and explore the potential of hybrid DCIM and ACIM approaches \cite{houshmand2022diana, chen2022hybrid, chen2023osa, yuan202434hybrid, guo202434hybrid}.

The paper is structured as follows:
\begin{itemize}
\item Section 2 details the fundamental concepts of CIM, including its basic architecture and operation principles.
\item Section 3 discusses Digital CIM (DCIM), examining its characteristics, advantages, and challenges.
\item Section 4 focuses on Analog CIM (ACIM), providing an in-depth analysis of when to use analog computing, various ACIM implementations, and a detailed look at charge-based ACIMs.
\item Section 5 reviews the designs of Hybrid CIMs and its advantages.
\item Section 6 concludes the paper, summarizing key points and discussing future directions in CIM research.
\end{itemize}
By examining the current state-of-the-art, we seek to offer insights into the challenges, opportunities, and potential future directions of the CIM technology. 

\section{Basic Concepts and Architecture of Compute-In-Memory (CIM)}

\begin{figure}[ht]
  \centerline{\includegraphics[width=0.5\linewidth]{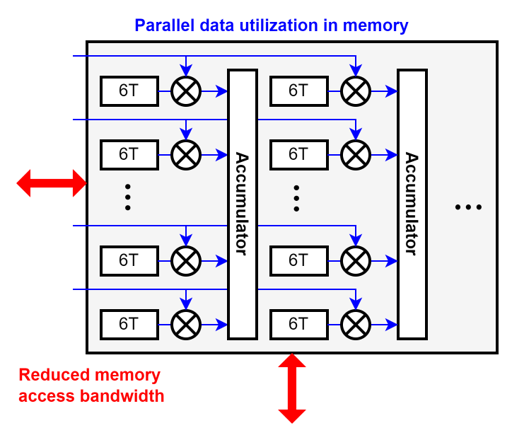}}
        \caption{Conceptional block diagram of a Compute-in-Memory Macro. }
        \label{CIMtop}
\end{figure}

Compute-In-Memory (CIM) is an innovative architecture that performs computations within or in close proximity to memory. This approach fundamentally rethinks the separation of memory and processor in traditional von Neumann architectures, aiming to significantly reduce the overhead associated with data movement. As shown in Fig.\ref{CIMtop}, the most crucial feature of CIM is its specialization in calculating the dot product of input vectors (IN) and weight vectors (W), which is the fundamental operation in neural networks.
A typical CIM macro consists of an SRAM cell array, multiplication circuits, accumulator circuits, input/output interfaces, and control logic. The SRAM cell array stores the weight vector (W), while the input vector (IN) is supplied through bitlines. This structure enables parallel execution of multiple weight-input multiplications, allowing for fast and efficient dot product calculations.

Here, the CIM operation with data mapping is explained. Mathematically, the dot product operation in CIM can be expressed as:
\begin{equation}
y = \sum_{i=1}^{n} IN_i \cdot W_i
\end{equation}
where $IN_i$ and $W_i$ are the $i$-th elements of the input and weight vectors, respectively, and $n$ is the vector length.
Specifically, the CIM architecture executes dot product calculations as follows:
\begin{enumerate}
\item Mapping of the Weight vector ($W$):
The weight vector $W$ is stored in the memory array, with each memory cell representing one bit of a weight value. For CIM macro with $n$ rows, each bit cell stores the weight value of $W_i$.
\item Mapping of the Input Vector ($IN$):
Each $IN_i$ is simultaneously applied to the corresponding bitline of the SRAM array. This means that $n$ inputs are processed in parallel, realizing the parallel processing from $i = 1$ to $n$ in the formula.

\item Multiplication Operation:
The weight $W_i$ stored in each SRAM cell is multiplied by the corresponding input value $IN_i$. In CIM architecture, this multiplication operation occurs simultaneously in each memory cell, using methods such as analog domain current multiplication or digital domain AND operations.

\item Addition (Accumulation) Operation:
The multiplication results $IN_i \cdot W_i$ are summed by the accumulator circuit. In ACIMs, this addition is performed along the wordline and is realized as a sum of currents or voltages. In DCIMs, this is realized through $n$-input digital adder trees.

\item Final Dot Product Output:
The result of the addition operation becomes the final output $y$. In ACIMs, the analog summed value is converted to a digital value through an analog-to-digital converter (ADC).
\end{enumerate}
This entire process occurs in parallel within the memory array, minimizing data movement compared to conventional architectures and enabling high-speed, low-power computations.

While the above explained the CIM operation for binary dot products, we can extend this to multi-bit representations \cite{shafiee2016isaac}. For multi-bit representation, we express $IN_i$ and $W_i$ as:
\begin{equation}
IN_i = \sum_{j=0}^{m-1} IN_{i,j} \cdot 2^j, \quad W_i = \sum_{k=0}^{p-1} W_{i,k} \cdot 2^k
\end{equation}
where $m$ and $p$ are the number of bits for input and weight, respectively.
The multi-bit dot product operation then becomes:
\begin{equation}
y = \sum_{j=0}^{m-1} \sum_{k=0}^{p-1} \left(\sum_{i=1}^{n} IN_{i,j} \cdot W_{i,k}\right) \cdot 2^{j+k}
\label{eq:dot-multi}
\end{equation}
Importantly, this equation shows that the core CIM circuit for computing dot products ($\sum_{i=1}^{n} IN_i \cdot W_i$) remains unchanged. The multi-bit dot product result $y$ is obtained by appropriately weighting (simply bit-shifting) and accumulating the results from the CIM circuit in the digital domain. This approach maintains the efficiency of the CIM architecture while extending its capability to handle multi-bit operations.

To summarize, the characteristic of CIM architecture is particularly effective for operations requiring large-scale matrix-vector multiplications, such as neural network forward propagation, self-attention mechanisms and convolution operations. For instance, calculations in fully connected layers or filter applications in convolutional layers can be efficiently executed using CIM.

\subsection{Digital and Analog CIMs}
\begin{figure}[ht]
  \centerline{\includegraphics[width=0.8\linewidth]{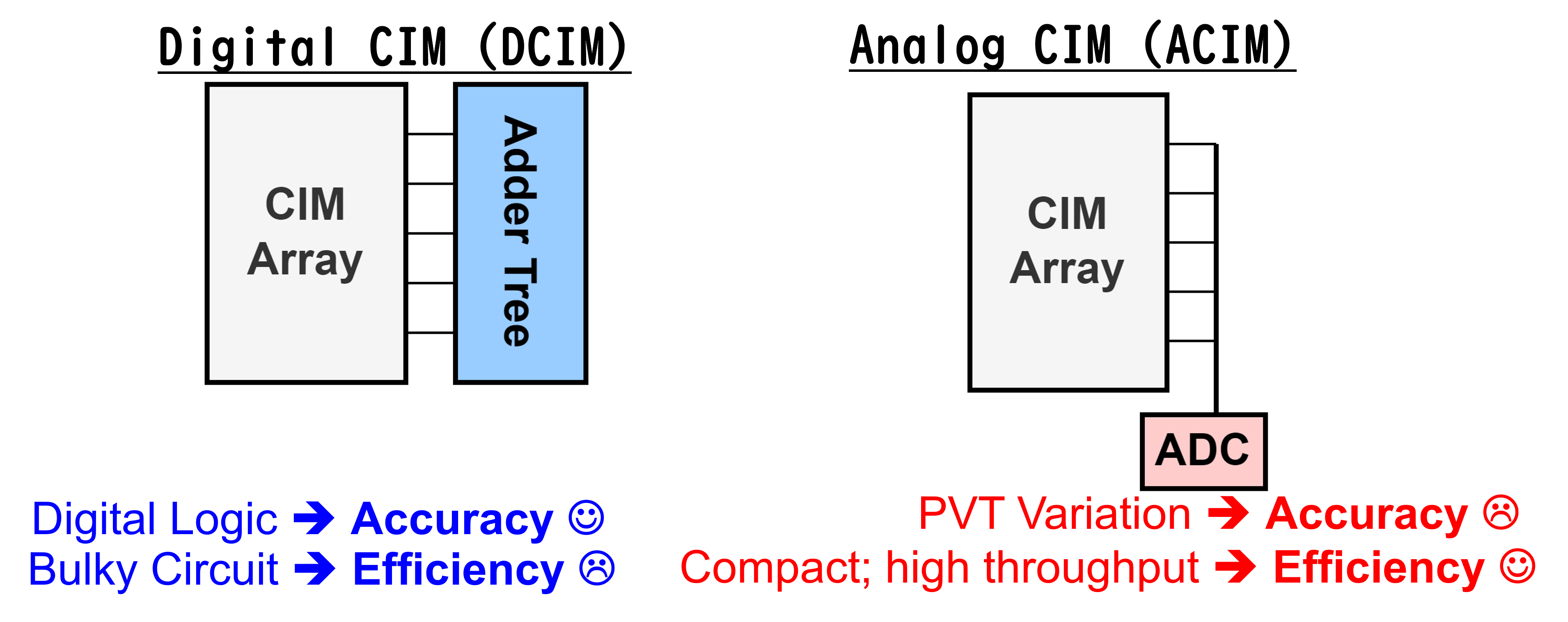}}
        \caption{Conceptional block diagram of a Compute-in-Memory Macro. }
        \label{dcim and acim}
\end{figure}

The implementation methods of CIM are primarily classified based on the design of the accumulator. This classification leads to three main categories: Digital CIM (DCIM), Analog CIM (ACIM), and Hybrid CIM, each with its own set of characteristics, advantages, and challenges.

Digital CIM (DCIM)~\cite{chih202116, zhang2024pacim,   tu2022trancim, tu2023multcim} employs digital adder trees to perform accumulation. This approach benefits from high precision, ease of design, and the ability to leverage existing digital circuit design techniques. However, DCIM tends to have lower area and power efficiency compared to its analog counterpart.

Analog CIM (ACIM)~\cite{gonugondla2020fundamental, valavi201964,  yao2023fully, yoshioka202434, yoshioka2024818, ando24ssdm,lee2021fully, jia2020programmable, jia202115, lee2022low, wu202228nmtime, dong202015current}, on the other hand, performs accumulation in the analog domain. Various methods have been proposed for ACIM, including operations in current~\cite{dong202015current}, charge ~\cite{valavi201964,  yao2023fully, yoshioka202434, yoshioka2024818, ando24ssdm,lee2021fully, jia2020programmable, jia202115, lee2022low}, or time domains~\cite{wu202228nmtime}. ACIM offers high power and area efficiency, as well as the potential for high-speed computation. Nevertheless, it faces challenges in terms of precision limitations, susceptibility to process variations, and scalability issues.

A third category, Hybrid CIM~\cite{houshmand2022diana, chen2022hybrid, chen2023osa}, aims to combine the strengths of both digital and analog designs. One example of this approach is processing the Most Significant Bits (MSB) digitally while handling the Least Significant Bits (LSB) in the analog domain. This hybrid method often allows for a better balance between precision and efficiency.

The selection among these implementation methods is determined by the required precision, power efficiency, area efficiency, and the characteristics of the target application. In the following chapters, we will delve deeper into DCIM and ACIM, exploring their features, advantages, and challenges in greater detail.

\section{Digital Compute-In-Memory (DCIM)}
Digital Compute-In-Memory (DCIM) is a variant of CIM architecture that implements the accumulator using digital circuits, specifically a Digital Adder Tree (DAT). DCIM aims to combine the advantages of digital circuit precision and design ease with the benefits of in-memory computation.

One of the primary strengths of DCIM is its high computational precision. The inherent nature of digital operations allows for maintaining error-free, high-accuracy calculations. 

Another significant of DCIM is its excellent scalability. As semiconductor manufacturing processes continue to advance, DCIM architectures can leverage these improvements to enhance both performance and energy efficiency. Recent reports have demonstrated successful DCIM implementations at process nodes as small as 3nm, indicating that DCIM can continue to progress in tandem with semiconductor technology advancements~\cite{fujiwara202434}. 

DCIM, despite its advantages, faces some design challenges. Integrating digital circuits within memory cells requires adherence to memory design rules rather than logic design rules. This constraint limits the use of automatic placement and routing tools post logic synthesis, necessitating manual circuit design by skilled layout designers. Consequently, this approach increases design complexity and potentially reduces productivity compared to conventional digital circuit design methodologies.

Power and area efficiency remain significant challenges for DCIM. Digital adders, essential components in DCIM, consume substantial power and occupy considerable area. 

\subsection{Review of digital CIM works}
\begin{figure}[ht]
  \centerline{\includegraphics[width=0.8\linewidth]{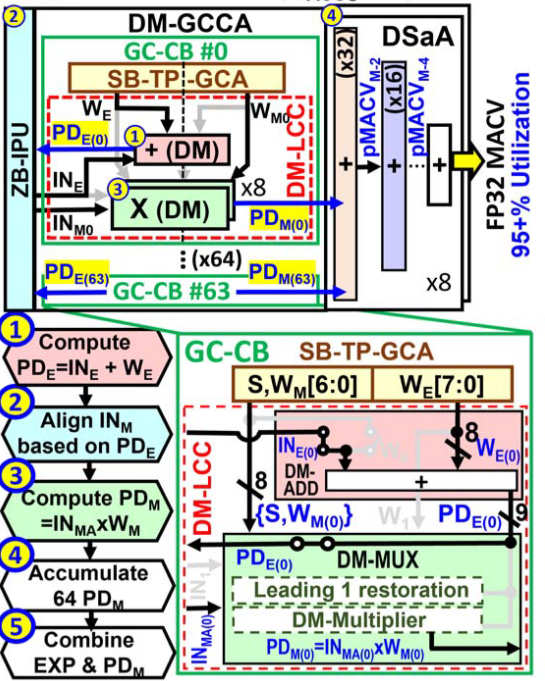}}
  \caption{Floating point (BF16) enabled DCIM macro (Figure adapted from \cite{khwa202434} \textcopyright IEEE)}
  \label{fig:float}
\end{figure}

Recent research has demonstrated the implementation of floating-point (BF16) operations in DCIM \cite{khwa202434}. This high precision makes DCIM particularly suitable for tasks demanding exceptional accuracy.
To enable floating-point operations, as illustrated in Fig.~\ref{fig:float}, exponent and mantissa computing units are integrated within the CIM cell. While floating-point support significantly expands the range of potential applications, it's important to note that the increased complexity of computational mechanisms within the bit cell may lead to concerns about reduced area efficiency.

\begin{figure}[ht]
  \centerline{\includegraphics[width=0.8\linewidth]{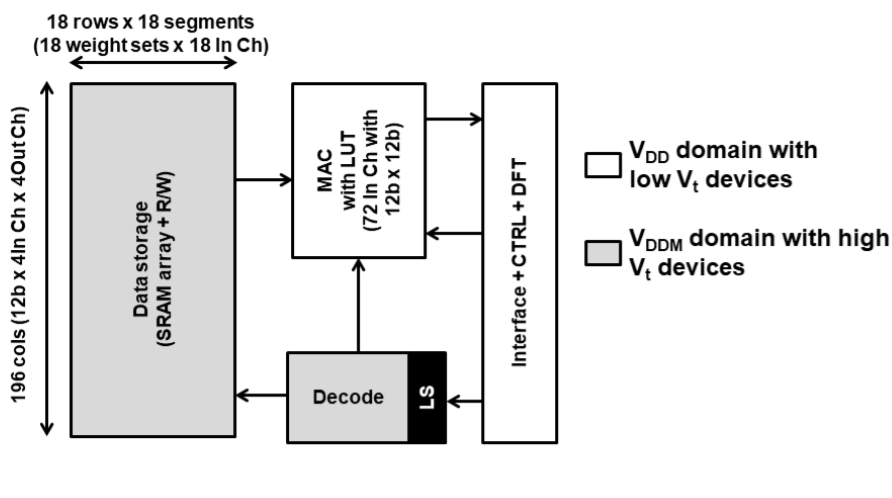}}
  \caption{INT12xINT12 DCIM macro (Figure adapted from  \cite{fujiwara202434} \textcopyright IEEE)}
  \label{fig:fujiwara_dcim}
\end{figure}

The DCIM macro presented at ISSCC'24 \cite{fujiwara202434} (Fig.~\ref{fig:fujiwara_dcim}) supports high-precision INT12xINT12 MAC operations. Notably, it separates data storage from computational circuitry, resembling a register and ALU unit architecture rather than traditional in-cell arithmetic units. This approach allows independent design of memory and arithmetic units, enhancing flexibility.

\begin{figure}[ht]
\centerline{\includegraphics[width=0.8\linewidth]{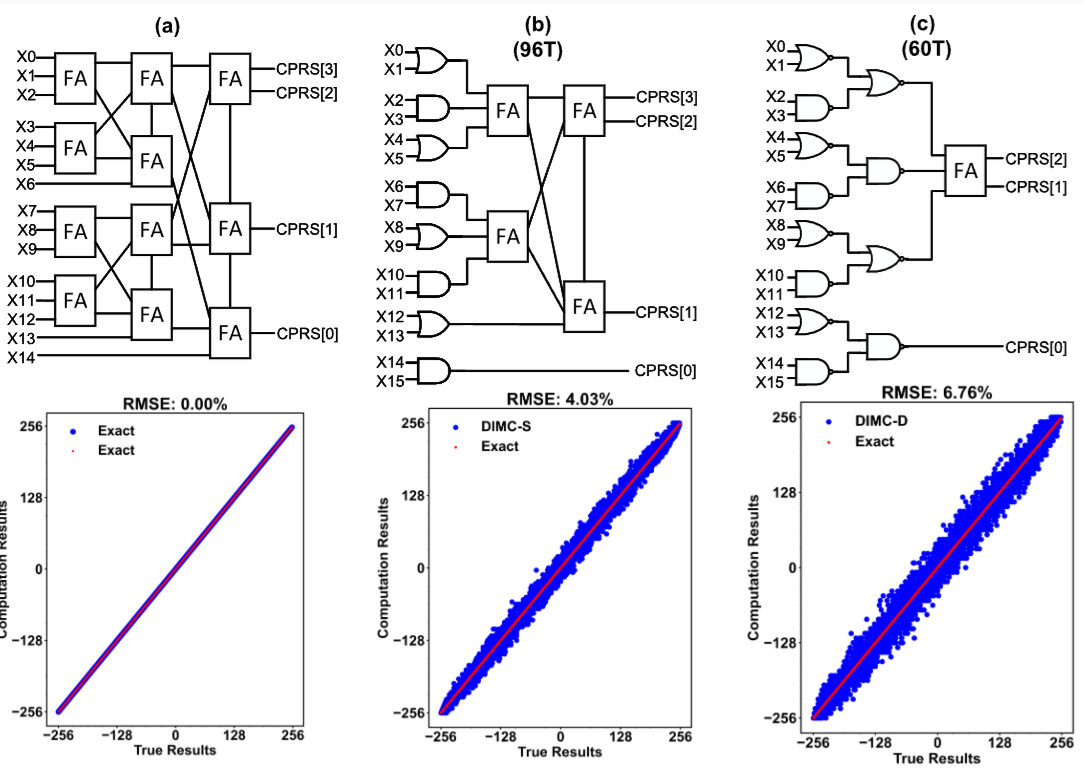}}
\caption{DCIM macro with approximate DAT circuits. (Figure adapted from \cite{lin2023dimca} \textcopyright IEEE)}
\label{fig:approx}
\end{figure}
To improve power and area efficiency in DCIMs, researchers have developed approximate digital CIM architectures. One approach approximates the DAT circuit \cite{lin2023dimca}, exploiting the noise tolerance of many DNN applications to trade accuracy for efficiency.
Fig.\ref{fig:approx} illustrates three DCIM designs with varying DAT approximation levels, where (a) is the typical error-free DAT implementation using a full array of full adders (FAs). \cite{lin2023dimca} shows that by  replacing the FAs for OR and AND gates for error-induced summation, DAT can trade accuracy for reduced transistor count. By accepting a RMSE of 6.7\%, \cite{lin2023dimca} achieves about 50\% less transistor count, achieving both power and area efficiency improvements.

\begin{figure}[ht]
\centerline{\includegraphics[width=0.8\linewidth]{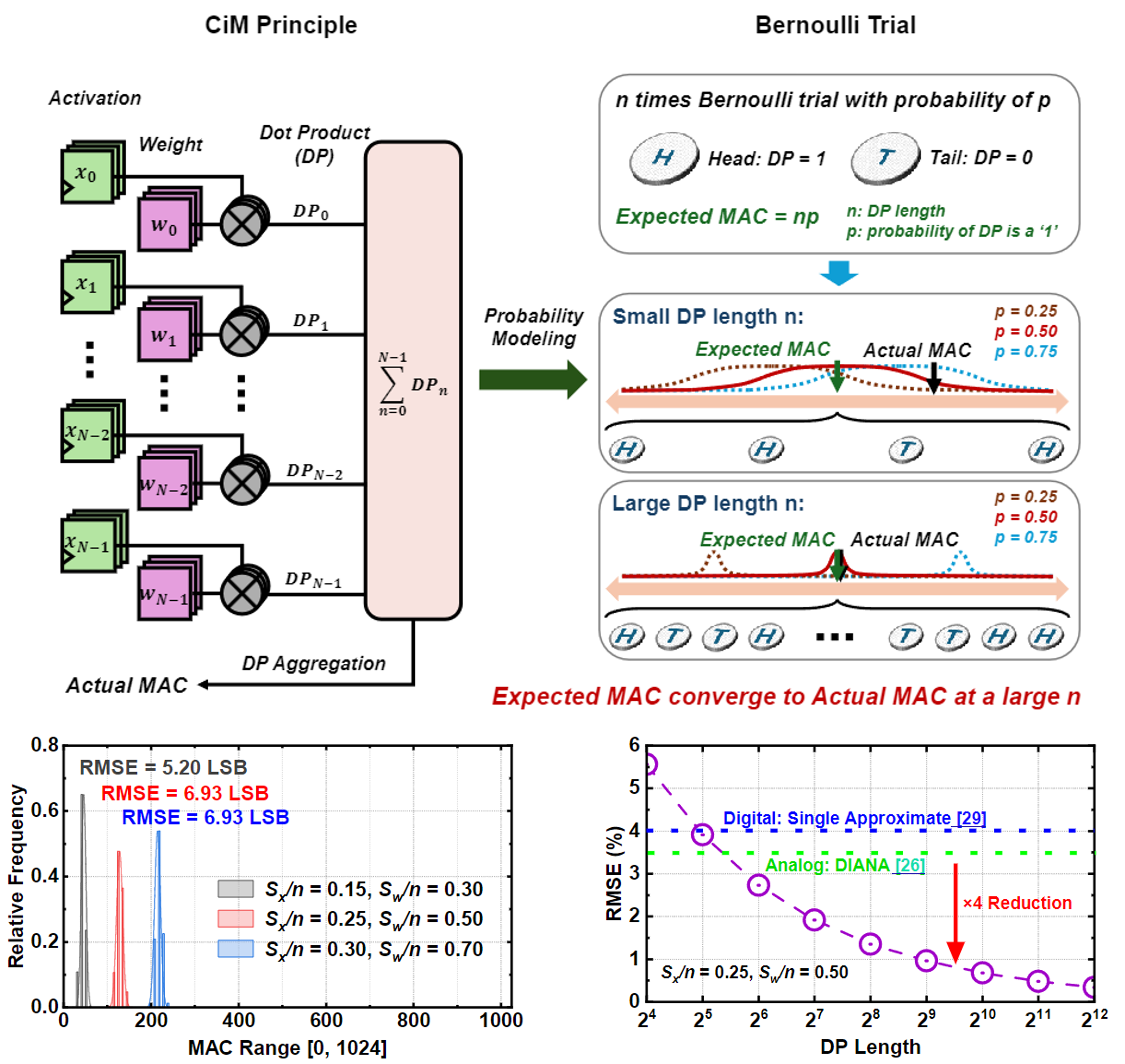}}
\caption{Approximating dot products with probabilistic computation. Figure adapted from \cite{zhang2024pacim} CC-BY-NC.}
\label{fig:picim}
\end{figure}

Another innovative design~\cite{zhang2024pacim} employs probabilistic computing techniques to approximate computations in DCIMs, as illustrated in Fig.~\ref{fig:picim}. This approach models dot product (DP) calculations as a series of Bernoulli trials, counting '1's in both input and weight vectors. Each element-wise multiplication is treated as a trial, with the probability of success determined by the proportion of '1's in the vectors. The dot product is then approximated by the number of successful trials in $n$ attempts, where $n$ is the vector length. This method exploits the law of large numbers, suggesting that for sufficiently large vector sizes, the approximation converges to the true value with high probability. The figure demonstrates this convergence, showing how the actual MAC result approaches the expected MAC as the DP length increases. 
The bottom graphs illustrate the error distribution and how RMSE decreases with increasing DP length, outperforming existing digital and analog approaches.

\section{Analog Compute-In-Memory (ACIM)}

Analog Compute-In-Memory (ACIM) is a CIM technique that leverages analog domain computations within the memory array to achieve high power and area efficiency~\cite{gonugondla2020fundamental, valavi201964,  yao2023fully, yoshioka202434, yoshioka2024818, ando24ssdm,lee2021fully, jia2020programmable, jia202115, lee2022low, wu202228nmtime, dong202015current, chen2024pico}. 
The fundamental architecture of ACIM eliminates the need for a DAT, which is typically required in DCIMs, by summing the multiplication results of IN and W in the analog domain. 
This approach follows a streamlined processing flow: (1) conversion of digital inputs to analog signals, (2) multiplication in the analog domain $(IN \times W)$, (3) summation in the analog domain $(\Sigma(IN \times W))$, and (4) conversion of analog results to digital via Analog-to-Digital Conversion (ADC). 

\subsection{When should we use analog computing?}

\begin{figure}[ht]
\centerline{\includegraphics[width=0.8\linewidth]{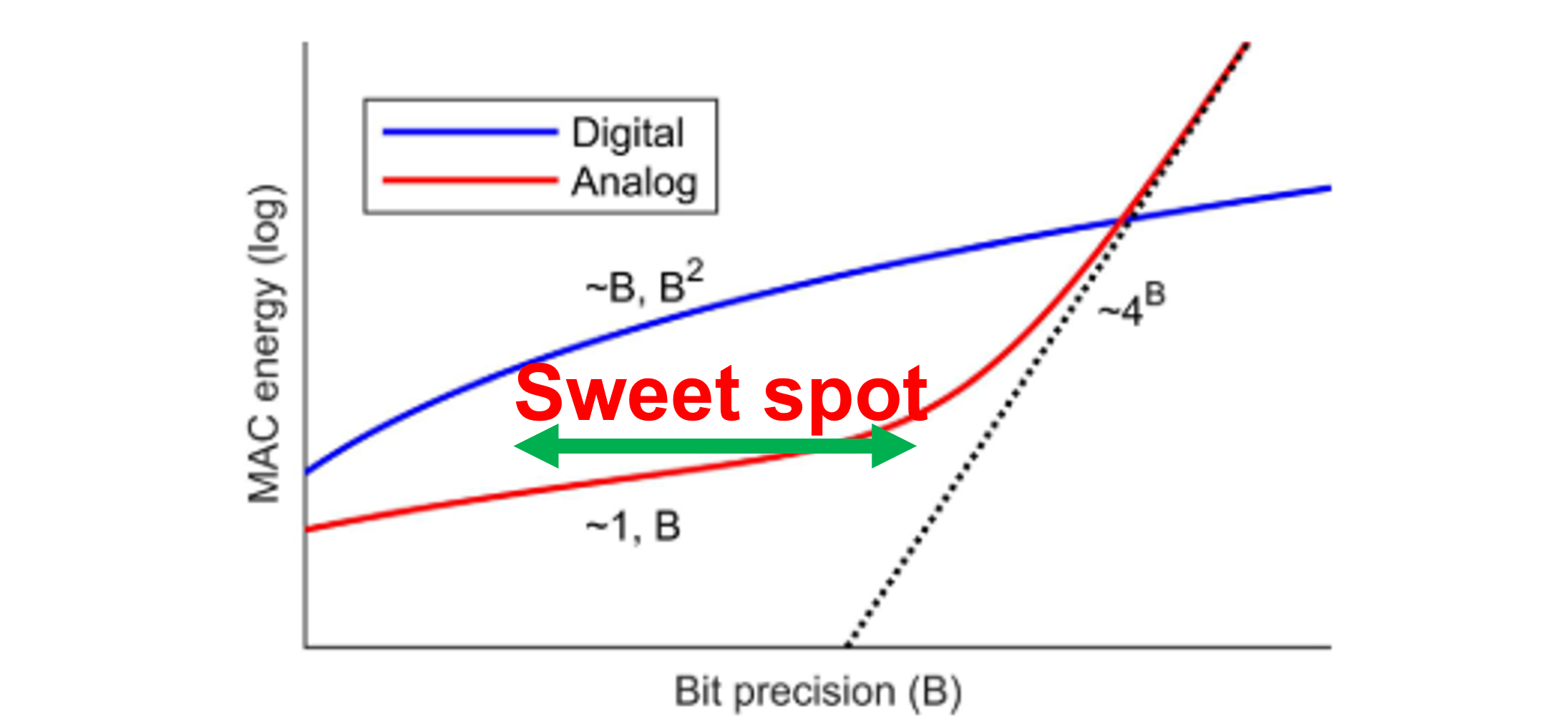}}
\caption{Digital and analog computing energy estimates (Figure adapted from \cite{murmann2020mixed} \textcopyright IEEE). The "Sweet spot" annotation is made by the author.}
\label{fig:murmann}
\end{figure}

The efficacy of analog computation is intricately tied to the required precision of the task at hand. Drawing from Murmann's research~\cite{murmann2020mixed} (Fig.~\ref{fig:murmann}), we can delineate the characteristics of analog and digital computing across different precision ranges:
\begin{itemize}
\item \textbf{Low Precision (1-2 bits):} Digital circuits demonstrate superior efficiency due to their inherent simplicity and speed advantages in this range.
\item \textbf{Medium Precision (3-8 bits):} Analog circuits excel, offering optimal power consumption and area efficiency without being significantly constrained by noise limitations. This range can be seen as the analog computing's "sweet spot".
\item \textbf{High Precision (10+ bits):} The efficiency of analog computation diminishes due to thermal noise constraints, making digital implementations more favorable.
\end{itemize}

Recent advancements in low-precision DNNs have demonstrated the viability of operating at bit-widths that fall within the optimal range for analog computing. For instance, cutting-edge research has shown successful quantization of CNNs to INT2 precision~\cite{choi2018pact}, while Vision Transformers (ViT), based on the Transformer architecture, have been effectively quantized to INT4~\cite{li2023repq}. These precision levels align remarkably well with the range where analog circuits exhibit superior efficiency, as illustrated in Figure \ref{fig:murmann}.

\subsection{Review of existing analog CIM implementations}

\begin{table}[ht]
\centering
\begin{tabular}{|p{1.5cm}|p{4cm}|p{4cm}|}
\cline{1-3}
 & \textbf{Pros} & \textbf{Cons} \\
\cline{1-3}
Current & Excellent area, power efficiency & Current susceptible to mismatch+PVT. \\
\cline{1-3}
Time & Excellent area, power efficiency & Delay susceptible to mismatch+PVT. \\
\cline{1-3}
Charge & Good area, power efficiency. Potential for high-precision. & Require area/energy hungry high-precision ADC. \\
\cline{1-3}
\end{tabular}
\caption{Comparison of Different ACIM Implementations}
\label{tab:acim_comparison}
\end{table}

As summarized in Table~\ref{tab:acim_comparison}, ACIM implementations can be broadly categorized into three main types: current-based, time-based, and charge-based. Each of these approaches offers unique advantages and challenges.

\textbf{Current-based ACIM} \cite{dong202015current} converts multiplication results into currents and performs accumulation via current integration. This approach enables highly area and power-efficient implementations, allowing for compact 7T cell designs. The simplicity of current conversion circuits, achievable with a single transistor, underlies this efficiency. However, the method faces a significant challenge: the inherent non-linearity of transistor currents. When using basic transconductance ($g_m$) cells, achieving the linearity required for accurate computations across PVTs becomes difficult due to the non-linear voltage-current relationship in MOSFETs. 

\textbf{Time-based ACIM} \cite{wu202228nmtime} converts multiplication results into delays, performing computations in the time domain. This approach utilizes delay-controlled cells for computation, which are essentially simple digital circuits, making them extremely compact and well-suited for implementation with deep-scaled transistors. Consequently, the design achieves high area and power efficiency. However, the primary challenge lies in the inherent non-linearity and PVT sensitivity of transistor delays.

\textbf{Charge-based ACIM}, the focus of our paper, converts multiplication results into charges and performs integration through charge redistribution. This approach leverages charge redistribution operations, extensively studied in successive approximation register (SAR) ADCs. The process can be implemented almost entirely with digital circuits, making it highly compatible with deep-scaled transistors and enabling high-speed, low-power execution. A key advantage is the use of metal-to-metal capacitors for charge storage, resulting in computing elements that are inherently robust against PVT variations. 
Recent research in SAR ADCs has demonstrated 12-bit linearity even in 3nm processes~\cite{lee20243nmsar}, showing the potential for high-precision operations. This technique has shown promising results in complex applications like ImageNet-class image recognition as well \cite{jia202115}.

\subsection{Charge-based ACIM operation}

\begin{figure}[ht]
\centerline{\includegraphics[width=1\linewidth]{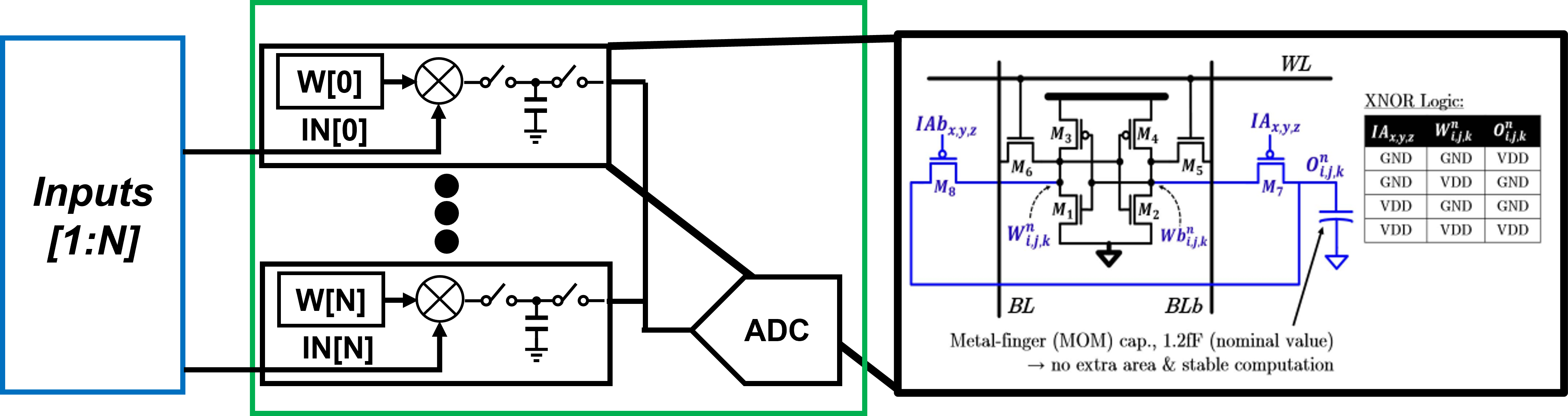}}
\caption{Structure and operation of the Multiplying Bit Cell (M-BC) in charge-based ACIM. Bit-cell figure adapted from \cite{valavi201964} \textcopyright IEEE}
\label{fig:charge}
\end{figure}

Charge-based ACIMs operate on a principle of charge manipulation and redistribution, as depicted in Figure \ref{fig:charge}. The process begins with recharging all capacitors to a reference voltage. Weights are set by adjusting charge quantities in corresponding capacitors based on SRAM cell states. Input data modulates bitline voltages, initiating charge redistribution between bitlines and capacitors—effectively performing multiplication. Row-wise charge integration yields matrix multiplication results, which are then digitized via ADC conversion.

The Multiplying Bit Cell (M-BC) structure~\cite{valavi201964} is a key innovation in this domain. It combines weight storage and multiplication within a compact layout, featuring a standard 6T SRAM cell enhanced with two pMOS transistors and a metal-oxide-metal (MOM) capacitor. This design efficiently executes XNOR operations between stored weights and input activations, accumulating results as charge on MOM capacitors.

MOM capacitors offer significant advantages: they're robust against PVT variations, maintaining consistent capacitance across diverse operating conditions. They also achieve high linearity up to 12-bit levels, contributing to the superior computational accuracy of charge-based ACIMs compared to current or time-domain approaches. Furthermore, MOM capacitors can be constructed entirely in metal layers above the 8T transistor structure, effectively eliminating area overhead and enhancing the M-BC's density and scalability.

\subsection{Quantifing ACIM computing precision with CSNR}
Compute SNR (CSNR) is a critical metric for evaluating the performance of ACIM systems, enabling fair comparisons with digital circuits and among different ACIM implementations. While traditional metrics such as INL and average ADC code noise have been used to assess ACIM performance, they are challenging to incorporate into software simulations and are not well-suited for comparing ACIMs. Here, we adopt the CSNR metric, as advocated in \cite{gonugondla2020fundamental, yoshioka2024818}, which provides a direct evaluation of the noise characteristics of analog circuits and can be easily integrated into software simulations.

Although \cite{gonugondla2020fundamental} initially introduced the theoretical foundation of CSNR, it did not provide a specific methodology for calculating CSNR from practical circuit measurement results. Here, we extend the analysis of \cite{gonugondla2020fundamental} by proposing a practical CSNR measurement method, enabling the effective quantification of analog noise and fair comparisons between high-precision and low-precision ACIMs as described in \cite{yoshioka2024818}. 

CSNR is conceptually similar to the SNR used in analog circuits and is defined as follows:
\begin{equation}
\mathrm{CSNR} = 20 \log_{10}\left(\frac{\mathrm{Signal}}{\mathrm{Noise}}\right)
\end{equation}
The noise model of ACIMs can be expressed using the ideal digital output $y_\mathrm{D}$, the ACIM output $y_\mathrm{A}$, the ADC quantization noise $Q$, and the analog noise $\gamma$:
\begin{equation}
y_\mathrm{A} = y_\mathrm{D} + Q + \gamma
\label{noise-eq}
\end{equation}
The analog noise $\gamma$ encompasses all non-idealities introduced by analog computation, with the primary sources being the nonlinearity of the MAC unit, thermal noise of transistors, device mismatch, and power supply noise. Intuitively, CSNR quantifies the ratio between the desired signal power and the power of the noise introduced by the analog computation process. A higher CSNR indicates that the analog computation is more accurate and closer to the ideal digital output, and vice versa.

To calculate CSNR from the computation results of a bit-serial ACIM, we consider an example where an 8-bit MAC output is obtained from 4-bit inputs and weights. In this scenario, the bit-serial ACIM performs 8 MAC operations, each involving a 1-bit×1-bit multiplication and accumulation. For each column of the bit-serial ACIM, a single 1b×1b MAC output $\mathrm{AMAC_{BS}}$ is mapped. $\mathrm{AMAC_{BS}}$ is computed using the input bit $\mathrm{IN}[i, N]$ (the $N$-th bit of the input) and the weight bit $\mathrm{Weight}[i, M]$ (the $M$-th bit of the weight):
\begin{equation}
\mathrm{AMAC_{BS}} = \sum \mathrm{IN}[i, N] \times \mathrm{Weight}[i, M]
\end{equation}
The 8 $\mathrm{AMAC_{BS}}$ values are then appropriately bit-shifted and accumulated to obtain the 8-bit MAC result $\mathrm{AMAC}$, which is intended to be equivalent to the digital 4b×4b MAC output ($\mathrm{DMAC}$):
\begin{equation}
\mathrm{AMAC} = \sum_{k=0}^{7} \mathrm{AMAC_{BS}}[k] \times 2^k
\end{equation}
where $\mathrm{AMAC_{BS}}[k]$ represents the $k$-th bit-serial computation result.

In general, bit-serial ACIMs are arranged in an array structure, allowing parallel processing of multiple columns. For example, a 4×4 array can simultaneously process the dot product of 4 input vectors and 4 weight vectors. In this case, the $\mathrm{AMAC}$ calculation described above is performed independently for each column, yielding 4 8-bit MAC results.
As $\mathrm{AMAC}$ contains the error shown in Eq. \ref{noise-eq}, CSNR can be calculated by treating the DMAC result ($\mathrm{DMAC}$) as the ideal signal and the absolute difference between $\mathrm{DMAC}$ and $\mathrm{AMAC}$ as noise:
\begin{equation}
\mathrm{CSNR} = 20 \log_{10}\left(\frac{\mathrm{DMAC}}{|\mathrm{DMAC} - \mathrm{AMAC}|}\right)
\end{equation}
However, when performing multiple trials, it is common to calculate CSNR using the average of the signal and noise powers (squared values):
\begin{equation}
\mathrm{CSNR} = 10 \log_{10}\left(\frac{\mathrm{mean}(\mathrm{DMAC}^2)}{\mathrm{mean}(|\mathrm{DMAC} - \mathrm{AMAC}|^2)}\right)
\end{equation}
Since a single computation result may not accurately capture the impact of Gaussian-distributed noise, it is recommended to use the average CSNR value obtained from 1,000 to 10,000 computation results. In this paper, we calculate CSNR using 10,000 trials, balancing the trade-off between convergence and computation time. 
\footnote{Open sourced in: \url{https://github.com/Keio-CSG/AnalogCIM-CSNR-Sim}}

\begin{figure}[ht]
\centerline{\includegraphics[width=0.8\linewidth]{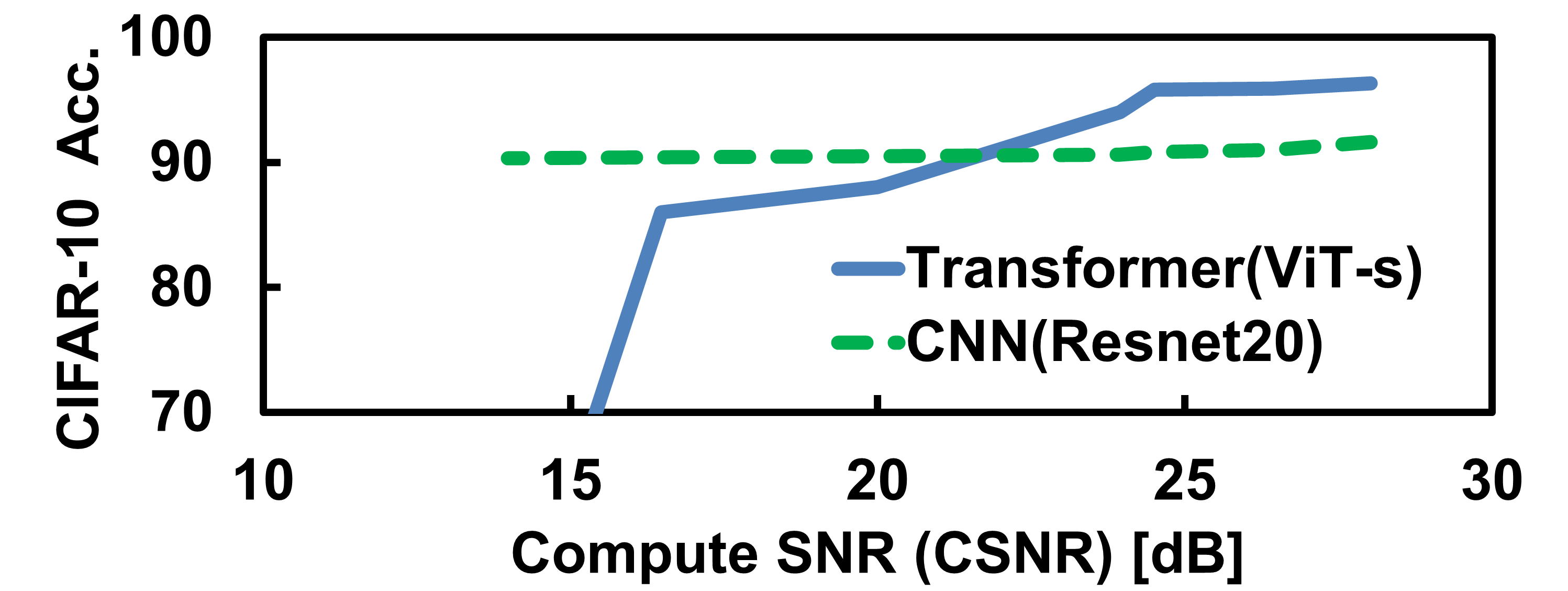}}
\caption{CIFAR10 accuracy changes when varying CSNR.}
\label{fig:cifar10}
\end{figure}

Using this metric, \cite{yoshioka2024818} conducted tests in PyTorch to simulate the addition of noise to DNN layers during inference, examining how CIFAR10 image recognition accuracy changes with CSNR variations (Fig.\ref{fig:cifar10}). The results showed that CNN (ResNet18) maintained nearly constant CIFAR10 recognition accuracy even as CSNR fluctuated between 15 and 30 dB. In contrast, Transformer (Vision Transformer) exhibited significant accuracy degradation with only slight CSNR deterioration.
These findings reveal that Transformers require higher precision computations compared to CNNs, while CNNs are well-suited for operation at low CSNR. Leveraging this characteristic, \cite{yoshioka2024818} optimizes ACIM at the algorithm level by operating CNNs at low CSNR to improve power efficiency, while running Transformers at high CSNR to enhance computational accuracy.

\subsection{Review of analog CIM works}

\begin{figure}[ht]
\centerline{\includegraphics[width=0.8\linewidth]{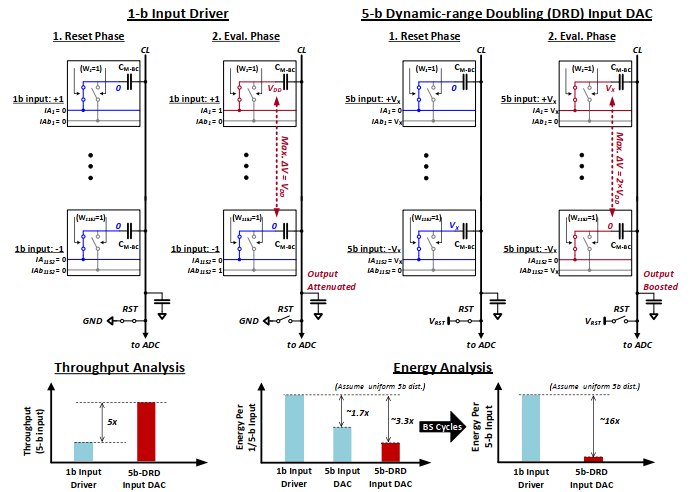}}
\caption{Bit-parallel operation in charge-based CIM. Figure adapted from \cite{lee2021fully} \textcopyright IEEE}
\label{fig:bp}
\end{figure}

\textbf{Bit-parallel analog CIMs:} While Section 2 assumed 1-bit input signals for CIM operations, charge-based CIM leverages the linear relationship between charge and voltage ($Q=CV$). This property allows for multi-bit analog voltage inputs, reducing the number of dot product operations related to IN by a factor of $n$ in equation \ref{eq:dot-multi}, thereby improving throughput and power efficiency.
Ref.~\cite{lee2021fully} proposes this "bit-parallel" operation that generates IN signals from 16 reference voltages, achieving a 5-bit DAC through Dynamic-range Doubling technology. This approach yields a 16-fold improvement in power efficiency compared to 1-bit bit-serial CIM operations. However, $n$-bit bit-parallel operation requires increased ADC resolution by $n$ bits to maintain the same readout precision as bit-serial operations, presenting a trade-off between efficiency and ADC complexity.

\begin{figure}[ht]
\centerline{\includegraphics[width=0.8\linewidth]{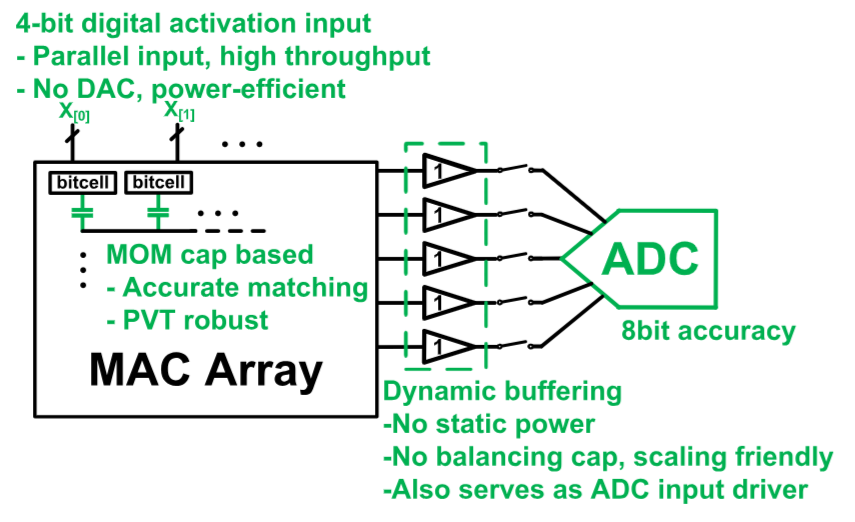}}
\caption{Multi-bit charge-based CIM. Figure adapted from \cite{yang20234oneshot} \textcopyright IEEE}
\label{fig:oneshot}
\end{figure}
\textbf{Multi-bit analog CIMs:} Conventional charge-based ACIMs place an ADC for each column, converting the accumulated analog charge per column and then applying weights in the digital domain. In contrast, multi-bit ACIMs (Fig.\ref{fig:oneshot}) use weighted capacitors to apply weights in the analog domain before accumulation, and then convert the result using a single ADC. By accumulating $n$ columns together in this manner, the number of ADCs required for computation can be reduced to $1/n$, promising significant efficiency improvements.
However, similar to the bit-parallel method, achieving high accuracy requires improving ADC precision. Additionally, the precision of the analog domain weighting must be sufficiently high, which may pose a design challenge. 

\begin{figure}[ht]
\centerline{\includegraphics[width=0.8\linewidth]{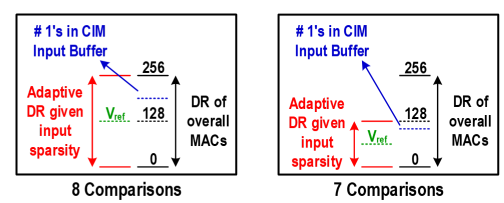}}
\caption{ACIM with sparsity sensing mechanism that dynamically adjusts ADC resolution by counting input '1's. Figure adapted from \cite{yao2023fully}. \textcopyright IEEE}
\label{fig:bp}
\end{figure}

\textbf{Sparcity analog CIMs:} In ACIM systems, ADC power consumption becomes a significant concern, necessitating effective power reduction strategies. Ref.~\cite{yao2023fully} achieves substantial ADC power reduction by leveraging input sparsity. The approach focuses on counting the number of '1's in the IN vector to estimate the maximum possible value of the dot product. For instance, in a 256-element dot product operation, if only 60 '1's are present in the input, the maximum dot product output is constrained to 60. This insight allows the system to deduce that the two most significant bits of an 8-bit SAR ADC result will be '0', effectively enabling the ADC to operate as a 6-bit converter. This dynamic adjustment effectively reducing ADC power consumption without compromising accuracy.

\begin{figure}[ht] \centerline{\includegraphics[width=0.8\linewidth]{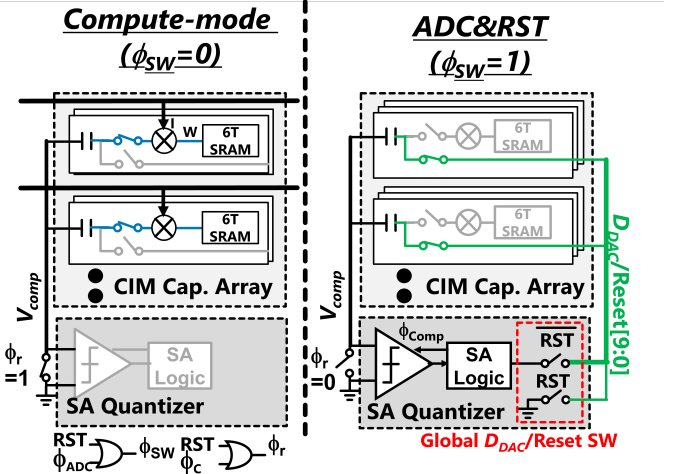}} \caption{Capacitor-reconfigured CIM structure integrating computation and the capacitor array of an SAR ADC. Figure adapted from \cite{yoshioka2024818} CC-BY-NC.} \label{fig:crcim} \end{figure}

\textbf{Transformer-capable analog CIMs:} Algorithms like Transformers require higher computational precision compared to CNNs. To achieve such precision with ACIM, the resolution of the ADC must be increased, which in SAR ADCs requires a larger capacitor array, leading to significant area overhead. In Fig.\ref{fig:crcim}\cite{yoshioka2024818}, a capacitor-reconfigured CIM structure is proposed, where the capacitor array used for SAR ADCs is integrated with computation, reusing the capacitors in the CIM bit-cell for AD conversion. This approach eliminates the area overhead associated with high-resolution ADCs, achieving 10-bit ADC conversion accuracy and providing a foundation for Transformer operations. However, improving precision beyond 10 bits is challenging due to thermal noise in analog circuits (Fig.\ref{fig:murmann}), necessitating alternative approaches for further accuracy enhancement.

\section{Hybrid CIMs}

\begin{figure}[ht]
\centerline{\includegraphics[width=0.8\linewidth]{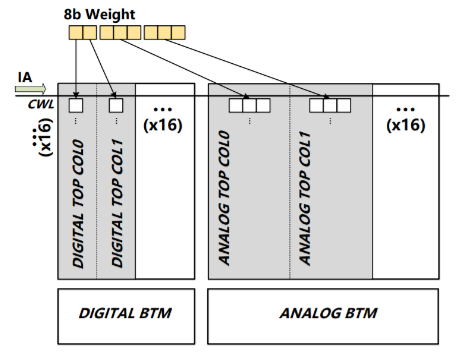}}
\caption{Hybrid CIM incorporating both ACIM and DCIM. Figure adapted from \cite{chen2022hybrid} \textcopyright IEEE.}
\label{fig:hybrid}
\end{figure}

\begin{figure}[ht]
\centerline{\includegraphics[width=0.8\linewidth]{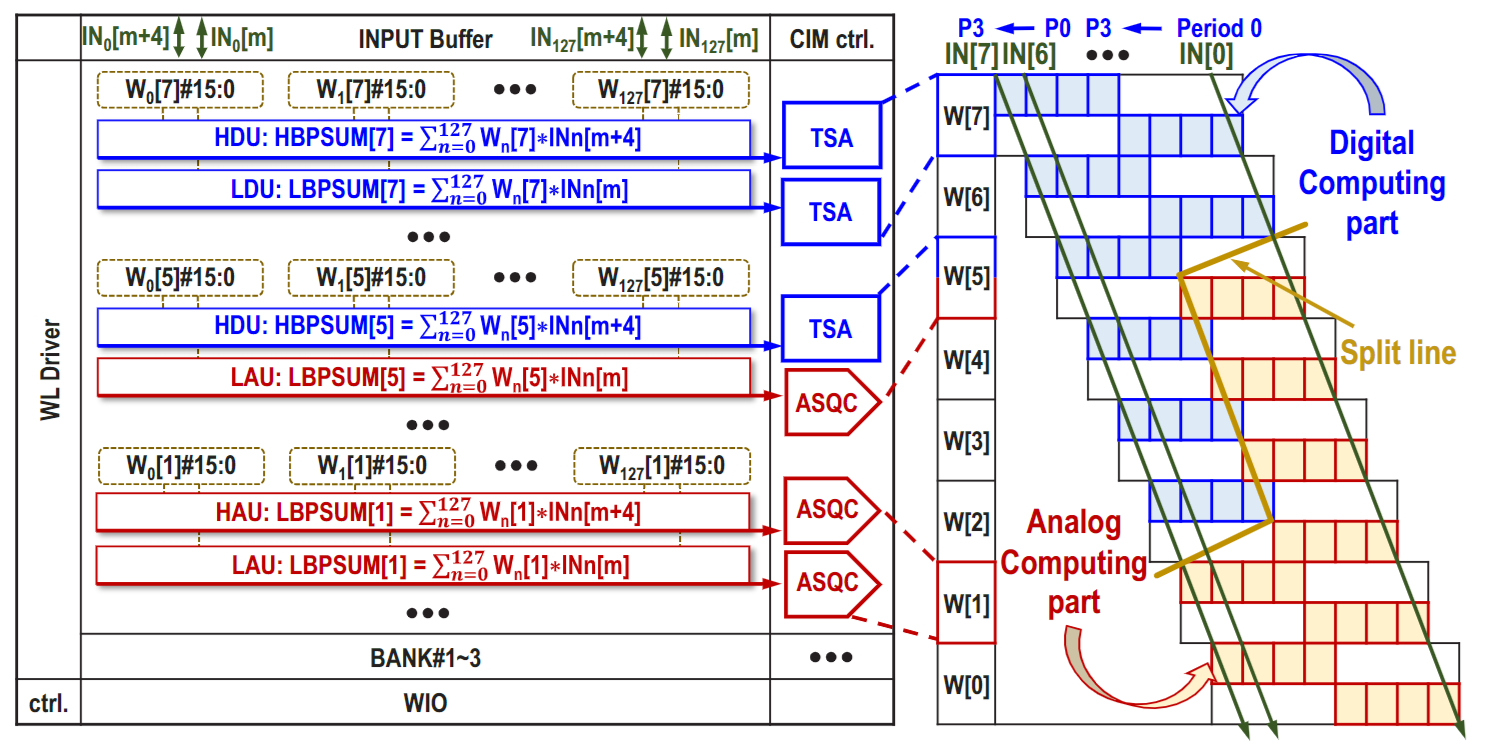}}
\caption{"Lightning" style Hybrid CIM. Figure adapted from \cite{guo202434} \textcopyright IEEE.}
\label{fig:lightning}
\end{figure}

So far, we've considered computation in either the analog or digital domain within the accumulator. A method combining both domains is referred to as Hybrid CIM. In \cite{chen2022hybrid}, the MSB portion of the dot product is computed using DCIM, while the LSB portion is handled by ACIM (Fig.\ref{fig:hybrid}). Since the MSB side is error-free, this approach maintains computation accuracy compared to ACIM while achieving lower power consumption than DCIM. Anpther recent hybrid CIM design \cite{guo202434} splits the DCIM and ACIM boundary in a "lightning" style, further balancing compute precision and energy efficiency (Fig.\ref{fig:lightning}).

From an ACIM perspective, hybridization offers enhanced MSB accuracy "for free." This allows for higher computational precision while exceeding the \textit{thermal} noise limits shown in Fig.\ref{fig:murmann}, unlocking the potential for further balance in power reduction and compute precision. As DNN algorithms become more complex and demand higher precision, hybrid computation will become a crucial approach. 

\begin{figure}[ht]
\centerline{\includegraphics[width=0.8\linewidth]{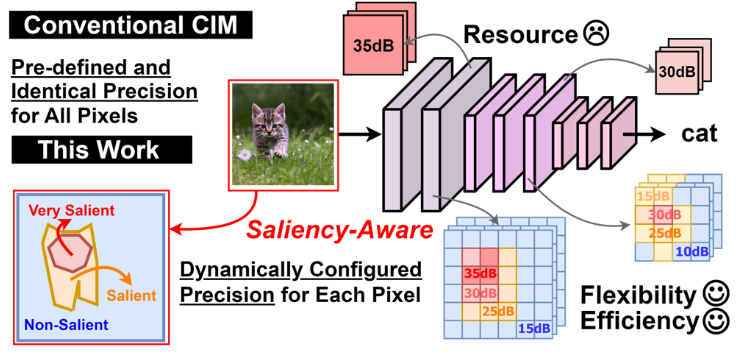}}
\caption{Data importance (Saliency). In image recognition, the cat is considered salient, significantly impacting algorithm outcomes, whereas the background is non-salient. Figure adapted from \cite{chen2023osaarxiv} CC-BY-NC.}
\label{fig:saliency}
\end{figure}

\begin{figure}[ht]
\centerline{\includegraphics[width=0.8\linewidth]{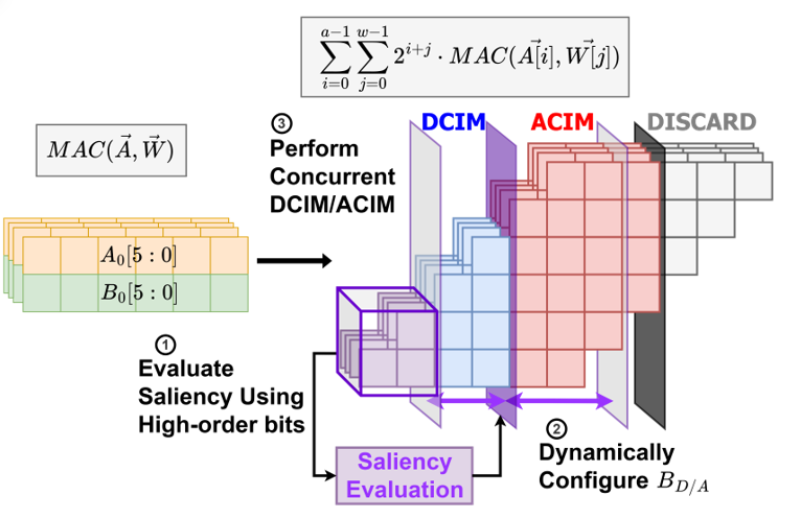}}
\caption{Saliency-based Hybrid CIM, dynamically adjusting the ratio of DCIM and ACIM based on saliency. Figure adapted from \cite{chen2023osaarxiv} CC-BY-NC.}
\label{fig:osa}
\end{figure}

To further enhance the power efficiency of Hybrid CIM, an innovative approach has been proposed that dynamically adjusts the balance between DCIM and ACIM based on the data’s importance (saliency) \cite{chen2023osa}. 
As shown in Fig.\ref{fig:saliency} and Fig.\ref{fig:osa}, the region of the cat is highly salient, where accuracy is critical. In such areas, DCIM is primarily used for its precision. Conversely, in less salient regions (e.g., the background), where computational precision is less crucial, ACIM is employed to reduce power consumption. 
Unlike conventional CIM sparsity that focuses on input data, the saliency-based computation determines algorithmic importance by evaluating MAC output values (in both FC and CNN layers) - higher output values indicate higher saliency and thus greater algorithmic significance.
To evaluate the saliency without any prior knowledge of the input or weights, our system uses the On-the-fly Saliency Evaluator (OSE). The key idea of OSE is to compute the s highest-order 1-bit MACs (s=2 in our implementation) using precise DCIM operations, and use these results to estimate the saliency of the entire MAC operation. 
These values are summed across channels to generate a saliency score. The OSE then uses this calculated saliency score to determine the Hybrid CIM MAC's digital-to-analog boundary ($B_{D/A}$) by comparing it against pre-trained thresholds. Based on this dynamic evaluation, our system can dynamically optimize the CIM's precision and power based on the saliency: high-saliency data are processed using more precise digital computation (DCIM), while low-saliency inputs use more analog computation (ACIM).

\section{System level implementations}
While our previous discussion focused primarily on macro-level implementations, this section addresses system-level implementations at the processor level. It's important to note that system-level power efficiency varies significantly depending on the algorithm size (which affects the number of external memory accesses), making direct comparisons challenging. Therefore, our analysis focuses more on the features achieved by these chips rather than purely comparing their efficiencies.

The research presented in \cite{tu202228nm} proposes a highly integrated CIM processor that achieves system-level efficiency of 30 TOPS/W. This implementation goes beyond basic CIM macros by incorporating pipelining and other processor-level optimizations. Additionally, it supports diverse computational precisions, including BF16 and INT8 formats.

Mediatek's reported work \cite{shih202420} presents a CIM processor designed for mobile accelerator applications, bringing this technology closer to commercial products. Their implementation is particularly interesting as it builds upon the CIM macro architecture from \cite{fujiwara202434}, refining it for practical processor applications. The processor is optimized for super-resolution applications and relatively small network sizes, allowing for an efficient system-level implementation without weight updates. They report approximately 5x improvement in system-level efficiency compared to processors using equivalent digital MAC architectures.

In these system-level implementations, DCIM has generally taken the lead over ACIM architectures. This advantage can be attributed to DCIM's deterministic operation, which makes it more manageable from a system perspective and simplifies compiler design. However, research into large-scale processor-level ACIM integration is also progressing, as demonstrated in \cite{jia202115}. While these ACIM implementations achieve good efficiency at the macro level, they require additional algorithmic considerations to address accuracy degradation, necessitating software-level compensation for ACIM characteristics. This challenge, however, presents an interesting research direction, as it could foster valuable collaboration between algorithm developers and analog circuit designers, potentially leading to significant advancements in the field.

\section{Conclusion and Future remarks}
In this paper, we have reviewed the key advancements in SRAM-based CIM circuits, focusing on both DCIM and ACIM implementations. While DCIM offers high computational precision and process scaling, ACIM presents benefits in power and area efficiency, particularly for applications requiring medium precision. 

Current primary limitations of CIM techniques can be categorized into accuracy concerns and weight-stationary characteristics. Regarding accuracy, DCIM demonstrates high performance, while ACIM can achieve improved accuracy through hybridization, making it applicable to CNN and ViT inference tasks at ImageNet scale. Also, advanced image processing tasks like semantic segmentation, which are fundamentally built on CNN layers, can theoretically be implemented using either CIM approach.

While LLMs share algorithmic similarities with ViT due to their transformer-based architecture, they face a fundamental challenge in implementing CIM's core principle. The basic concept of CIM aims to minimize data movement by keeping weights stationary within memory. However, as models grow larger, maintaining stationary weights within on-chip SRAM becomes problematic. Since the gigantic weight matrices must be read from off-chip memory, this often requires weight data to be rewritten between layers. When this occurs, CIM's distinctive advantage over conventional PEs diminishes, as frequent weight updates effectively nullify the benefits of in-memory computation. This may represent a critical bottleneck in scaling CIM applications to larger models.

\section*{Acknowledgements}
This research was supported in part by the JST CREST JPMJCR21D2 and JSPS Kakenhi 23H00467.

\section*{References}
\bibliographystyle{IEEEtran} 
\bibliography{reference}

\begin{thebibliography}{10}
\providecommand{\url}[1]{#1}
\csname url@samestyle\endcsname
\providecommand{\newblock}{\relax}
\providecommand{\bibinfo}[2]{#2}
\providecommand{\BIBentrySTDinterwordspacing}{\spaceskip=0pt\relax}
\providecommand{\BIBentryALTinterwordstretchfactor}{4}
\providecommand{\BIBentryALTinterwordspacing}{\spaceskip=\fontdimen2\font plus
\BIBentryALTinterwordstretchfactor\fontdimen3\font minus \fontdimen4\font\relax}
\providecommand{\BIBforeignlanguage}[2]{{%
\expandafter\ifx\csname l@#1\endcsname\relax
\typeout{** WARNING: IEEEtran.bst: No hyphenation pattern has been}%
\typeout{** loaded for the language `#1'. Using the pattern for}%
\typeout{** the default language instead.}%
\else
\language=\csname l@#1\endcsname
\fi
#2}}
\providecommand{\BIBdecl}{\relax}
\BIBdecl

\bibitem{krizhevsky2012imagenet}
A.~Krizhevsky, I.~Sutskever, and G.~E. Hinton, ``Imagenet classification with deep convolutional neural networks,'' \emph{Advances in neural information processing systems}, vol.~25, 2012.

\bibitem{he2016resnet}
K.~He, X.~Zhang, S.~Ren, and J.~Sun, ``Deep residual learning for image recognition,'' in \emph{CVPR}, 2016, pp. 770--778.

\bibitem{vaswani2017attention}
A.~Vaswani, ``Attention is all you need,'' \emph{Advances in Neural Information Processing Systems}, 2017.

\bibitem{he2015delving}
K.~He, X.~Zhang, S.~Ren, and J.~Sun, ``Delving deep into rectifiers: Surpassing human-level performance on imagenet classification,'' in \emph{Proceedings of the IEEE international conference on computer vision}, 2015, pp. 1026--1034.

\bibitem{brown2020language}
T.~B. Brown, ``Language models are few-shot learners,'' \emph{arXiv preprint arXiv:2005.14165}, 2020.

\bibitem{horowitz20141}
M.~Horowitz, ``1.1 computing's energy problem (and what we can do about it),'' in \emph{2014 IEEE international solid-state circuits conference digest of technical papers (ISSCC)}.\hskip 1em plus 0.5em minus 0.4em\relax IEEE, 2014, pp. 10--14.

\bibitem{sebastian2020memory}
A.~Sebastian, M.~Le~Gallo, R.~Khaddam-Aljameh, and E.~Eleftheriou, ``Memory devices and applications for in-memory computing,'' \emph{Nature nanotechnology}, vol.~15, no.~7, pp. 529--544, 2020.

\bibitem{gonugondla2020fundamental}
S.~K. Gonugondla, C.~Sakr, H.~Dbouk, and N.~R. Shanbhag, ``{Fundamental limits on the precision of in-memory architectures},'' in \emph{Proceedings of the 39th International Conference on Computer-Aided Design}, 2020, pp. 1--9.

\bibitem{verma2019memory}
N.~Verma, H.~Jia, H.~Valavi, Y.~Tang, M.~Ozatay, L.-Y. Chen, B.~Zhang, and P.~Deaville, ``In-memory computing: Advances and prospects,'' \emph{IEEE Solid-State Circuits Magazine}, vol.~11, no.~3, pp. 43--55, 2019.

\bibitem{valavi201964}
H.~Valavi, P.~J. Ramadge, E.~Nestler, and N.~Verma, ``{A 64-tile 2.4-Mb in-memory-computing CNN accelerator employing charge-domain compute},'' \emph{IEEE Journal of Solid-State Circuits}, vol.~54, no.~6, pp. 1789--1799, 2019.

\bibitem{yao2023fully}
C.-Y. Yao, T.-Y. Wu, H.-C. Liang, Y.-K. Chen, and T.-T. Liu, ``{A Fully Bit-Flexible Computation in Memory Macro Using Multi-Functional Computing Bit Cell and Embedded Input Sparsity Sensing},'' \emph{IEEE Journal of Solid-State Circuits}, 2023.

\bibitem{yoshioka202434}
K.~Yoshioka, ``{A 818-4094TOPS/W Capacitor-Reconfigured CIM Macro for Unified Acceleration of CNNs and Transformers},'' in \emph{International Solid-State Circuits Conference (ISSCC)}, vol.~67.\hskip 1em plus 0.5em minus 0.4em\relax IEEE, 2024, pp. 574--576.

\bibitem{yoshioka2024818}
------, ``{A 818--4094 TOPS/W Capacitor-Reconfigured Analog CIM for Unified Acceleration of CNNs and Transformers},'' \emph{IEEE Journal of Solid-State Circuits}, 2024.

\bibitem{ando24ssdm}
S.~Ando, S.~Miyagi, Y.~Chen, W.~Zhang, and K.~Yoshioka, ``{A Saliency-Aware Analog Computing-In-Memory Macro with SAR-Embedded Saliency Detection Technique},'' in \emph{2024 SSDM}.\hskip 1em plus 0.5em minus 0.4em\relax JJAP, 2024, pp. 1--2.

\bibitem{jia2020programmable}
H.~Jia, H.~Valavi, Y.~Tang, J.~Zhang, and N.~Verma, ``{A programmable heterogeneous microprocessor based on bit-scalable in-memory computing},'' \emph{IEEE Journal of Solid-State Circuits}, vol.~55, no.~9, pp. 2609--2621, 2020.

\bibitem{jia202115}
H.~Jia, M.~Ozatay, Y.~Tang, H.~Valavi, R.~Pathak, J.~Lee, and N.~Verma, ``{A Programmable Neural-Network Inference Accelerator based on Scalable In-Memory Computing},'' in \emph{International Solid-State Circuits Conference (ISSCC)}, vol.~64.\hskip 1em plus 0.5em minus 0.4em\relax IEEE, 2021, pp. 236--238.

\bibitem{lee2021fully}
J.~Lee, H.~Valavi, Y.~Tang, and N.~Verma, ``{Fully row/column-parallel in-memory computing SRAM macro employing capacitor-based mixed-signal computation with 5-b inputs},'' in \emph{2021 Symposium on VLSI Circuits}.\hskip 1em plus 0.5em minus 0.4em\relax IEEE, 2021, pp. 1--2.

\bibitem{lee2022low}
K.~Lee, J.~Kim, and J.~Park, ``{Low-Cost 7T-SRAM Compute-In-Memory Design based on Bit-Line Charge-Sharing based Analog-To-Digital Conversion},'' in \emph{Proceedings of the 41st IEEE/ACM International Conference on Computer-Aided Design}, 2022, pp. 1--8.

\bibitem{chen2024pico}
Z.~Chen, Z.~Wen, W.~Wan, A.~R. Pakala, Y.~Zou, W.-C. Wei, Z.~Li, Y.~Chen, and K.~Yang, ``{PICO-RAM: A PVT-Insensitive Analog Compute-In-Memory SRAM Macro With In Situ Multi-Bit Charge Computing and 6T Thin-Cell-Compatible Layout},'' \emph{IEEE Journal of Solid-State Circuits}, 2024.

\bibitem{wu202228nmtime}
P.-C. Wu, J.-W. Su, Y.-L. Chung, L.-Y. Hong, J.-S. Ren, F.-C. Chang, Y.~Wu, H.-Y. Chen, C.-H. Lin, H.-M. Hsiao \emph{et~al.}, ``{A 28nm 1Mb time-domain computing-in-memory 6T-SRAM macro with a 6.6 ns latency, 1241GOPS and 37.01 TOPS/W for 8b-MAC operations for edge-AI devices},'' in \emph{International Solid-State Circuits Conference (ISSCC)}, vol.~65.\hskip 1em plus 0.5em minus 0.4em\relax IEEE, 2022, pp. 1--3.

\bibitem{dong202015current}
Q.~Dong, M.~E. Sinangil, B.~Erbagci, D.~Sun, W.-S. Khwa, H.-J. Liao, Y.~Wang, and J.~Chang, ``{A 351TOPS/W and 372.4 GOPS compute-in-memory SRAM macro in 7nm FinFET CMOS for machine-learning applications},'' in \emph{International Solid-State Circuits Conference-(ISSCC)}.\hskip 1em plus 0.5em minus 0.4em\relax IEEE, 2020, pp. 242--244.

\bibitem{houshmand2022diana}
P.~Houshmand, G.~M. Sarda, V.~Jain, K.~Ueyoshi, I.~A. Papistas, M.~Shi, Q.~Zheng, D.~Bhattacharjee, A.~Mallik, P.~Debacker \emph{et~al.}, ``{Diana: An end-to-end hybrid digital and analog neural network soc for the edge},'' \emph{IEEE Journal of Solid-State Circuits}, vol.~58, no.~1, pp. 203--215, 2022.

\bibitem{chen2022hybrid}
J.~Chen, T.~Xiong, and X.~Si, ``{A Charge-Digital Hybrid Compute-In-Memory Macro with full precision 8-bit Multiply-Accumulation for Edge Computing Devices},'' in \emph{International Symposium on Embedded Multicore/Many-core Systems-on-Chip (MCSoC)}.\hskip 1em plus 0.5em minus 0.4em\relax IEEE, 2022, pp. 153--158.

\bibitem{chen2023osa}
Y.-C. Chen, S.~Ando, D.~Fujiki, S.~Takamaeda-Yamazaki, and K.~Yoshioka, ``{OSA-HCIM: On-The-Fly Saliency-Aware Hybrid SRAM CIM with Dynamic Precision Configuration},'' \emph{IEEE ASP-DAC}, 2024.

\bibitem{yuan202434hybrid}
Y.~Yuan, Y.~Yang, X.~Wang, X.~Li, C.~Ma, Q.~Chen, M.~Tang, X.~Wei, Z.~Hou, J.~Zhu \emph{et~al.}, ``{34.6 A 28nm 72.12 TFLOPS/W Hybrid-Domain Outer-Product Based Floating-Point SRAM Computing-in-Memory Macro with Logarithm Bit-Width Residual ADC},'' in \emph{2024 IEEE International Solid-State Circuits Conference (ISSCC)}, vol.~67.\hskip 1em plus 0.5em minus 0.4em\relax IEEE, 2024, pp. 576--578.

\bibitem{guo202434hybrid}
A.~Guo, X.~Chen, F.~Dong, J.~Chen, Z.~Yuan, X.~Hu, Y.~Zhang, J.~Zhang, Y.~Tang, Z.~Zhang \emph{et~al.}, ``{34.3 A 22nm 64kb Lightning-Like Hybrid Computing-in-Memory Macro with a Compressed Adder Tree and Analog-Storage Quantizers for Transformer and CNNs},'' in \emph{2024 IEEE International Solid-State Circuits Conference (ISSCC)}, vol.~67.\hskip 1em plus 0.5em minus 0.4em\relax IEEE, 2024, pp. 570--572.

\bibitem{chih202116}
Y.-D. Chih, P.-H. Lee, H.~Fujiwara, Y.-C. Shih, C.-F. Lee, R.~Naous, Y.-L. Chen, C.-P. Lo, C.-H. Lu, H.~Mori \emph{et~al.}, ``{An 89TOPS/W and 16.3 TOPS/mm 2 all-digital SRAM-based full-precision compute-in memory macro in 22nm for machine-learning edge applications},'' in \emph{International Solid-State Circuits Conference (ISSCC)}, vol.~64.\hskip 1em plus 0.5em minus 0.4em\relax IEEE, 2021, pp. 252--254.

\bibitem{zhang2024pacim}
W.~Zhang, S.~Ando, Y.-C. Chen, S.~Miyagi, S.~Takamaeda-Yamazaki, and K.~Yoshioka, ``Pacim: A sparsity-centric hybrid compute-in-memory architecture via probabilistic approximation,'' \emph{arXiv preprint arXiv:2408.16246}, 2024.

\bibitem{tu2022trancim}
F.~Tu, Z.~Wu, Y.~Wang, L.~Liang, L.~Liu, Y.~Ding, L.~Liu, S.~Wei, Y.~Xie, and S.~Yin, ``{TranCIM: Full-digital bitline-transpose CIM-based sparse transformer accelerator with pipeline/parallel reconfigurable modes},'' \emph{IEEE Journal of Solid-State Circuits}, 2022.

\bibitem{tu2023multcim}
F.~Tu, Z.~Wu, Y.~Wang, W.~Wu, L.~Liu, Y.~Hu, S.~Wei, and S.~Yin, ``{MulTCIM: Digital Computing-in-Memory-Based Multimodal Transformer Accelerator With Attention-Token-Bit Hybrid Sparsity},'' \emph{IEEE Journal of Solid-State Circuits}, 2023.

\bibitem{fujiwara202434}
H.~Fujiwara, H.~Mori, W.-C. Zhao, K.~Khare, C.-E. Lee, X.~Peng, V.~Joshi, C.-K. Chuang, S.-H. Hsu, T.~Hashizume \emph{et~al.}, ``{34.4 A 3nm, 32.5 TOPS/W, 55.0 TOPS/mm 2 and 3.78 Mb/mm 2 Fully-Digital Compute-in-Memory Macro Supporting INT12$\times$ INT12 with a Parallel-MAC Architecture and Foundry 6T-SRAM Bit Cell},'' in \emph{2024 IEEE International Solid-State Circuits Conference (ISSCC)}, vol.~67.\hskip 1em plus 0.5em minus 0.4em\relax IEEE, 2024, pp. 572--574.

\bibitem{khwa202434}
W.-S. Khwa, P.-C. Wu, J.-J. Wu, J.-W. Su, H.-Y. Chen, Z.-E. Ke, T.-C. Chiu, J.-M. Hsu, C.-Y. Cheng, Y.-C. Chen \emph{et~al.}, ``{34.2 A 16nm 96Kb Integer/Floating-Point Dual-Mode-Gain-Cell-Computing-in-Memory Macro Achieving 73.3-163.3 TOPS/W and 33.2-91.2 TFLOPS/W for AI-Edge Devices},'' in \emph{2024 IEEE International Solid-State Circuits Conference (ISSCC)}, vol.~67.\hskip 1em plus 0.5em minus 0.4em\relax IEEE, 2024, pp. 568--570.

\bibitem{shih202420digital}
M.-E. Shih, S.-W. Hsieh, P.-Y. Tsai, M.-H. Lin, P.-K. Tsung, E.-J. Chang, J.~Liang, S.-H. Chang, C.-L. Huang, Y.-Y. Nian \emph{et~al.}, ``{20.1 NVE: A 3nm 23.2 TOPS/W 12b-Digital-CIM-Based Neural Engine for High-Resolution Visual-Quality Enhancement on Smart Devices},'' in \emph{2024 IEEE International Solid-State Circuits Conference (ISSCC)}, vol.~67.\hskip 1em plus 0.5em minus 0.4em\relax IEEE, 2024, pp. 360--362.

\bibitem{yoshioka24ssdm}
K.~Yoshioka, S.~Ando, S.~Miyagi, Y.~Chen, and W.~Zhang, ``{Towards Efficient and Precise Analog Compute-in-Memory Circuits},'' in \emph{2024 SSDM}.\hskip 1em plus 0.5em minus 0.4em\relax JJAP, 2024, pp. 1--2.

\bibitem{shafiee2016isaac}
A.~Shafiee, A.~Nag, N.~Muralimanohar, R.~Balasubramonian, J.~P. Strachan, M.~Hu, R.~S. Williams, and V.~Srikumar, ``{ISAAC: A convolutional neural network accelerator with in-situ analog arithmetic in crossbars},'' \emph{ACM SIGARCH Computer Architecture News}, vol.~44, no.~3, pp. 14--26, 2016.

\bibitem{lin2023dimca}
C.-T. Lin, D.~Wang, B.~Zhang, G.~K. Chen, P.~C. Knag, R.~K. Krishnamurthy, and M.~Seok, ``Dimca: An area-efficient digital in-memory computing macro featuring approximate arithmetic hardware in 28 nm,'' \emph{IEEE Journal of Solid-State Circuits}, 2023.

\bibitem{murmann2020mixed}
B.~Murmann, ``{Mixed-signal computing for deep neural network inference},'' \emph{IEEE Transactions on Very Large Scale Integration (VLSI) Systems}, vol.~29, no.~1, pp. 3--13, 2020.

\bibitem{choi2018pact}
J.~Choi, Z.~Wang, S.~Venkataramani, P.~I.-J. Chuang, V.~Srinivasan, and K.~Gopalakrishnan, ``Pact: Parameterized clipping activation for quantized neural networks,'' \emph{arXiv preprint arXiv:1805.06085}, 2018.

\bibitem{li2023repq}
Z.~Li, J.~Xiao, L.~Yang, and Q.~Gu, ``Repq-vit: Scale reparameterization for post-training quantization of vision transformers,'' in \emph{Proceedings of the IEEE/CVF International Conference on Computer Vision}, 2023, pp. 17\,227--17\,236.

\bibitem{lee20243nmsar}
S.~Lee, J.~Park, J.~Park, S.~Lee, J.~Lee, Y.~Cho, M.~Choi, and J.~Shin, ``{3.9 A 1.2 V High-Voltage-Tolerant Bootstrapped Analog Sampler in 12-bit SAR ADC Using 3nm GAA’s 0.7 V Thin-Gate-Oxide Transistor},'' in \emph{2024 IEEE International Solid-State Circuits Conference (ISSCC)}, vol.~67.\hskip 1em plus 0.5em minus 0.4em\relax IEEE, 2024, pp. 70--72.

\bibitem{yang20234oneshot}
X.~Yang and N.~Sun, ``{A 4-bit mixed-signal MAC macro with one-shot ADC conversion},'' \emph{IEEE Journal of Solid-State Circuits}, vol.~58, no.~9, pp. 2648--2658, 2023.

\bibitem{guo202434}
A.~Guo, X.~Chen, F.~Dong, J.~Chen, Z.~Yuan, X.~Hu, Y.~Zhang, J.~Zhang, Y.~Tang, Z.~Zhang \emph{et~al.}, ``{A 22nm 64kb Lightning-Like Hybrid Computing-in-Memory Macro with a Compressed Adder Tree and Analog-Storage Quantizers for Transformer and CNNs},'' in \emph{International Solid-State Circuits Conference (ISSCC)}, vol.~67.\hskip 1em plus 0.5em minus 0.4em\relax IEEE, 2024, pp. 570--572.

\bibitem{chen2023osaarxiv}
Y.-C. Chen, S.~Ando, D.~Fujiki, S.~Takamaeda-Yamazaki, and K.~Yoshioka, ``{OSA-HCIM: On-The-Fly Saliency-Aware Hybrid SRAM CIM with Dynamic Precision Configuration},'' \emph{arXiv preprint arXiv:2308.15040}, 2023.

\bibitem{tu202228nm}
F.~Tu, Y.~Wang, Z.~Wu, L.~Liang, Y.~Ding, B.~Kim, L.~Liu, S.~Wei, Y.~Xie, and S.~Yin, ``{A 28nm 29.2 TFLOPS/W BF16 and 36.5 TOPS/W INT8 reconfigurable digital CIM processor with unified FP/INT pipeline and bitwise in-memory booth multiplication for cloud deep learning acceleration},'' in \emph{2022 IEEE International Solid-State Circuits Conference (ISSCC)}, vol.~65.\hskip 1em plus 0.5em minus 0.4em\relax IEEE, 2022, pp. 1--3.

\bibitem{shih202420}
M.-E. Shih, S.-W. Hsieh, P.-Y. Tsai, M.-H. Lin, P.-K. Tsung, E.-J. Chang, J.~Liang, S.-H. Chang, C.-L. Huang, Y.-Y. Nian \emph{et~al.}, ``{20.1 NVE: A 3nm 23.2 TOPS/W 12b-Digital-CIM-Based Neural Engine for High-Resolution Visual-Quality Enhancement on Smart Devices},'' in \emph{2024 IEEE International Solid-State Circuits Conference (ISSCC)}, vol.~67.\hskip 1em plus 0.5em minus 0.4em\relax IEEE, 2024, pp. 360--362.

\end{thebibliography}

\end{document}